\documentstyle [12pt,epsf]{article}

\input{epsf}
\begin{document}
\begin{titlepage}
\begin{center}

 \vspace{-0.7in}

{\large \bf Stochastic Quantization for Complex Actions}\\
 \vspace{.3in}{\large\em G. Menezes\,\footnotemark[1] and N.~F.~Svaiter \footnotemark[2]}\\
\vspace{.3in}
Centro Brasileiro de Pesquisas F\'{\i}sicas\,-CBPF,\\
 Rua Dr. Xavier Sigaud 150,\\
 Rio de Janeiro, RJ, 22290-180, Brazil \\

\subsection*{\\Abstract}
\end{center}
\baselineskip .18in

We use the stochastic quantization method to study systems with
complex valued path integral weights. We assume a Langevin
equation with a memory kernel and Einstein's relations with
colored noise. The equilibrium solution of this non-Markovian
Langevin equation is analyzed. We show that for a large class of
elliptic non-Hermitian operators acting on scalar functions on
Euclidean space, which define different models in quantum field
theory, converges to an equilibrium state in the asymptotic limit
of the Markov parameter $\tau\rightarrow\infty$. Moreover, as we
expected, we obtain the Schwinger functions of the theory.

\vspace{0,3cm}
 PACS numbers: 03.70+k, 04.62.+v

\footnotetext[1]{e-mail:\,\,gsm@cbpf.br}
\footnotetext[2]{e-mail:\,\,nfuxsvai@cbpf.br}

\end{titlepage}
\newpage\baselineskip .20
in
\section{Introduction}
\quad

The program of stochastic quantization, first proposed by Parisi
and Wu \cite{parisi}, and the stochastic regularization was
carried out for generic fields defined in flat, Euclidean
manifolds. A brief introduction to stochastic quantization can be
found in Refs. \cite{sakita} \cite{ii} \cite{namiki1}, and a
complete review is given in Ref. \cite{damre}. Recently Menezes
and Svaiter \cite{menezes3} implemented the stochastic
quantization in the theory of self-interacting scalar fields in a
static Riemannian manifold and also a manifold with a event
horizon, namely, the Einstein and the Rindler manifold. First,
these authors solved a Langevin equation for the mode coefficients
of the field, then they exhibit the two-point function at the
one-loop level. It was shown that it diverges and to regularize
the theory they used a covariant stochastic regularization. The
presence of the Markov parameter as an extra dimension allows the
authors to implement a regularization scheme, which preserves all
the symmetries of the theory under study. It is clear that the
stochastic quantization program can be implemented without
problems, if it is possible to perform the Wick rotation,
obtaining a real Euclidean action.

The picture that emerges from the above discussion is that the
implementation of the stochastic quantization in curved background
is related to the following fact. For static manifold, it is
possible to perform a Wick rotation, i.e., analytically extend the
pseudo-Riemannian manifold to the Riemannian domain without
problem. Nevertheless, for non-static curved manifolds we have to
extend the formalism beyond the Euclidean signature, i.e., to
formulate the stochastic quantization in pseudo-Riemannian
manifold, not in the Riemannian space (as in the original
Euclidean space) as was originally formulated. See for example the
discussion presented by Huffel and Rumpf \cite{huff} and Gozzi
\cite{gozzi}. In the first of these papers the authors proposed a
modification of the original Parisi-Wu scheme, introducing a
complex drift term in the Langevin equation, to implement the
stochastic quantization in Minkowski spacetime. Gozzi studied the
spectrum of the non-self-adjoint Fokker-Planck Hamiltonian to
justify this program. See also the papers \cite{huff2}
\cite{call}. Of course, these situations are special cases of
ordinary Euclidean formulation for systems with complex actions.

The main difference between the implementation of the stochastic
quantization in Minkowski spacetime and in Euclidean space is the
fact that in the latter case the approach to the equilibrium state
is a stationary solution of the Focker-Planck equation. In the
Minkowski formulation, the Hamiltonian is non-Hermitian and the
eigenvalues of such Hamiltonian are in general complex. The real
part of such eigenvalues are important to the asymptotic behavior
at large Markov time, and the approach to the equilibrium is
achieved only if we can show its positive semi-definiteness. The
crucial question is: what happens if the Langevin equation
describes diffusion around complex action? Some authors claim that
it is possible to obtain meaningful results out of Langevin
equation diffusion processes around complex action. Parisi
\cite{con4} and Klauder and Peterson \cite{con5} investigated the
complex Langevin equation, where some numerical simulations in
one-dimensional systems was presented. See also the papers
\cite{con6} \cite{con7}. We would also like to mention the
approach developed by Okamoto et al. \cite{okamoto} where the role
of the kernel in the complex Langevin equation was studied.

We would like to remark that there are many examples where
Euclidean action is complex. The simplest case is the stochastic
quantization in Minkowski spacetime, as we discussed. Other
situations are systems, as for example $QCD$, with non-vanishing
chemical potential at finite temperature; for $SU(N)$ theories
with $N>2$, the fermion determinant becomes complex and also the
effective action. Complex terms can also appear in the Langevin
equation for fermions, but a suitable kernel can circumvent this
problem \cite{f1} \cite{f2} \cite{f3}. Another case that deserves
our attention is the stochastic quantization of topological field
theories. One of the peculiar features within these kind of
theories is the appearance of a factor $i$ in front of the
topological actions in Euclidean space. In these topological
theories, the path integral measure weighing remain to be
$e^{iS}$, even after the Wick rotation. An attempt to use a
Markovian Langevin equation with a white noise to quantize the
theory fails since the Langevin equation will not tend to any
equilibrium at large Markov parameter. In the literature there are
different proposes to solve the above mentioned problem. In a pure
topological Chern-Simons theory, Ferrari et al. \cite{con8}
introduced a non-trivial kernel in the Langevin equation. Other
approach was developed by Menezes and Svaiter \cite{menezes1}.
These authors proved that, using a non-Markovian Langevin equation
with a colored random noise, it is possible to obtain convergence
towards equilibrium even with an imaginary Chern-Simons
coefficient. An interesting application of this method can be
found on Ref. \cite{menezes2}, where a Langevin equation with a
memory kernel was introduced in order to obtain the Schwinger
functions for the self-interacting scalar model. In conclusion,
although several alternative methods have been proposed to deal
with interesting physical systems where the Euclidean action is
complex \cite{salcedo} \cite{gaus} \cite{gaus1} \cite{sd}, these
methods do not suggest any general way of solving the particular
difficulties that arise in each situation. Here, we wish to report
progress in the stochastic quantization of theories with imaginary
action, introducing a memory kernel.

It is the purpose of the present paper to use the method of the
stochastic quantization to study systems with complex valued path
integral weights. We assume a Langevin equation with a memory
kernel and Einstein's relations with colored noise \cite{zz}. We
show that for a large class of elliptic non-Hermitean operators
which define different models in quantum field theory converges in
the asymptotic limit of the Markov parameter
$\tau\rightarrow\infty$, and we obtain the Schwinger functions of
the theory. In section II, we briefly discuss the Parisi-Wu
stochastic quantization for the case of free scalar field. In
section III we implement the stochastic quantization for scalar
field with complex action using a non-Markovian Langevin equation.
Conclusions are given in section IV. In this paper we use
$\hbar=c=k_{B}=1$.

\section{Stochastic quantization for the
free scalar field theory: the Euclidean case}

In this section, we give a brief survey of the stochastic
quantization. This technique in flat spacetime with trivial
topology can be summarized by the following steps. First, starting
from a field defined in Minkowski spacetime, after analytic
continuation to imaginary time, the Euclidean counterpart, i.e.,
the field defined in an Euclidean space, is obtained. Second, it
is introduced a monotonically crescent Markov parameter, called in
the literature ``fictitious time" and also a random noise field
$\eta(\tau,x)$, which simulates the coupling between the classical
system and a heat reservoir. It is assumed that the fields defined
at the beginning in a $d$-dimensional Euclidean space also depends
on the Markov parameter, therefore the field and a random noise
field are defined in a $(d+1)$-dimensional manifold. One starts
with the system out of equilibrium at an arbitrary initial state.
It is then forced into equilibrium assuming that its evolution is
governed by a Markovian Langevin equation with a white random
noise field \cite{kac} \cite{z} \cite{kubo}. In fact, this
evolution is described by a process which is stationary, Gaussian
and Markovian. Finally, the $n$-point correlation functions of the
theory in the $(d+1)$-dimensional space are defined by performing
averages over the random noise field with a Gaussian distribution,
that is, performing the stochastic averages
$\langle\,\varphi(\tau_{1},x_{1})
\varphi(\tau_{2},x_{2})...\varphi(\tau_{n},x_{n})\,\rangle_{\eta}$.
The $n$-point Schwinger functions of the Euclidean $d$-dimensional
theory are obtained  evaluating these $n$-point stochastic
averages $\langle\,\varphi(\tau_{1},x_{1})
\varphi(\tau_{2},x_{2})...\varphi(\tau_{n},x_{n})\,\rangle_{\eta}$
when the Markov parameter goes to infinity
$(\tau\rightarrow\infty)$, and the equilibrium is reached. This
can be proved in different ways for the particular case of
Euclidean scalar field theory. One can use, for instance, the
Fokker-Planck equation \cite{fp} \cite{ili} associated with the
equations describing the stochastic dynamic of the system. A
diagrammatical technique \cite{grimus} has also been used to prove
such equivalence.

After this brief digression, let us consider a free neutral scalar
field. The Euclidean action that usually describes such theory is
\begin{equation}
S_{0}[\varphi]=\int d^{d}x\, \left(\frac{1}{2}
(\partial\varphi)^{2}+\frac{1}{2}
m_{0}^{2}\,\varphi^{2}(x)\right). \label{9}
\end{equation}

The simplest starting point of the stochastic quantization to
obtain the Euclidean field theory is a Markovian Langevin
equation. Assume a flat Euclidean $d$-dimensional manifold, where
we are choosing periodic boundary conditions for a scalar field
and also a random noise. In other words, they are defined in a
$d$-torus $\Omega\equiv\,T\,^d$. To implement the stochastic
quantization we supplement the scalar field $\varphi(x)$ and the
random noise $\eta(x)$ with an extra coordinate $\tau$, the Markov
parameter, such that $\varphi(x)\rightarrow \varphi(\tau,x)$ and
$\eta(x)\rightarrow \eta(\tau,x)$. Therefore, the fields and the
random noise are defined in a domain: $T\,^{d}\times R\,^{(+)}$.
Let us consider that this dynamical system is out of equilibrium,
being described by the following equation of evolution:
\begin{equation}
\frac{\partial}{\partial\tau}\varphi(\tau,x)
=-\frac{\delta\,S_0}{\delta\,\varphi(x)}|_{\varphi(x)=\varphi(\tau,\,x)}+
\eta(\tau,x), \label{23}
\end{equation}
where $\tau$ is a Markov parameter, $\eta(\tau,x)$ is a random
noise field and $S_0$ is the usual free Euclidean action defined
in Eq. (\ref{9}). For a free scalar field, the Langevin equation
reads
\begin{equation}
\frac{\partial}{\partial\tau}\varphi(\tau,x)
=-(-\Delta+m^{2}_{0}\,)\varphi(\tau,x)+ \eta(\tau,x),
 \label{24}
\end{equation}
where $\Delta$ is the $d$-dimensional Laplace operator. The Eq.
(\ref{24}) describes a Ornstein-Uhlenbeck process and we are
assuming the Einstein relations, that is:
\begin{equation}
\langle\,\eta(\tau,x)\,\rangle_{\eta}=0, \label{28}
\end{equation}
and for the two-point correlation function associated with the
random noise field
\begin{equation}
\langle\, \eta(\tau,x)\,\eta(\tau',x')\,
\rangle_{\eta}\,=2\delta(\tau-\tau')\,\delta^{d}(x-x'), \label{29}
\end{equation}
where $\langle\,...\rangle_{\eta}$ means stochastic averages. The
above equation defines a delta-correlated random process. In a
generic way, the stochastic average for any functional of
$\varphi$ given by $F[\varphi\,]$ is defined by
\begin{equation}
\langle\,F[\varphi\,]\,\rangle_{\eta}=
\frac{\int\,[d\eta]F[\varphi\,]\exp\biggl[-\frac{1}{4} \int d^{d}x
\int d\tau\,\eta^{2}(\tau,x)\bigg]}
{\int\,[d\eta]\exp\biggl[-\frac{1}{4} \int d^{d}x \int
d\tau\,\eta^{2}(\tau,x)\bigg]}. \label{36}
\end{equation}
Let us define the retarded Green function for the diffusion problem
that we call $G(\tau-\tau',x-x')$. The retarded Green function
satisfies $G(\tau-\tau',x-x')=0$ if $\tau-\tau'<0$ and also
\begin{equation}
\Biggl[\frac{\partial}{\partial\tau}+(-\Delta_{x}+m^{2}_{0}\,)
\Bigg]G(\tau-\tau',x-x')=\delta^{d}(x-x')\delta(\tau-\tau').
\label{25}
\end{equation}
Using the retarded Green function and the initial condition
$\varphi(\tau,x)|_{\tau=0}=0$, the solution for Eq. (\ref{24})
reads
\begin{equation}
\varphi(\tau,x)=\int_{0}^{\tau}d\tau'\int_{\Omega}d^{d}x'\,
G(\tau-\tau',x-x')\eta(\tau',x'). \label{26}
\end{equation}
Let us define the Fourier transforms for the field  and the noise
given by $\varphi(\tau,k)$ and $\eta(\tau,k)$. We have respectively
\begin{equation}
\varphi(\tau,k)=\frac{1}{(2\pi)^\frac{d}{2}} \int\,d^{d}x\,
e^{-ikx}\,\varphi(\tau,x), \label{33}
\end{equation}
and
\begin{equation}
\eta(\tau,k)=\frac{1}{(2\pi)^\frac{d}{2}} \int\,d^{d}x\,
e^{-ikx}\,\eta(\tau,x). \label{34}
\end{equation}
Substituting Eq. (\ref{33}) in Eq. (\ref{9}), the free action for
the scalar field in the $(d+1)$-dimensional space writing in terms
of the Fourier coefficients reads
\begin{equation}
S_{0}[\varphi(k)]\,|_{\varphi(k)=\varphi(\tau,\,k)}=
\frac{1}{2}\int\,d^{d}k\,\varphi(\tau,k)(k^{2}+m_{0}^{2})\varphi(\tau,k).
\label{35}
\end{equation}
Substituting Eq. (\ref{33}) and Eq. (\ref{34}) in Eq. (\ref{24})
we have that each Fourier coefficient satisfies a Langevin
equation given by
\begin{equation}
\frac{\partial}{\partial\tau}\varphi(\tau,k)
=-(k^{2}+m^{2}_{0})\varphi(\tau,k)+ \eta(\tau,k). \label{36}
\end{equation}
In the Langevin equation the particle is subject to a fluctuating
force (representing a stochastic environment), where its average
properties are presumed to be known and also the friction force.
Note that the "friction coefficient" in the Eq. (\ref{36}) is
given by $(k^{2}+m^{2}_{0})$.

The solution for Eq. (\ref{36}) reads
\begin{equation}
\varphi(\tau,k)=\exp\left(-(k^{2}+m_{0}^{2})\tau\right)\varphi(0,k)+
\int_{0}^{\tau}d\tau'\exp\left(-(k^{2}+m_{0}^{2})(\tau-\tau')\right)
\eta(\tau',k). \label{37}
\end{equation}
Using the Einstein relation, we get that the Fourier coefficients
for the random noise satisfies
\begin{equation}
\langle\,\eta(\tau,k)\,\rangle_{\eta}=0 \label{38}
\end{equation}
and
\begin{equation}
\langle\,\eta(\tau,k)\eta(\tau',k')\,\rangle
_{\eta}=2\,\delta(\tau-\tau')\,\delta^{d}(k+k'). \label{39}
\end{equation}
It is possible to show that
$\langle\,\varphi(\tau,k)\varphi(\tau',k')\,\rangle_{\eta}|_{\tau=\tau'}\equiv
D(k,k';\tau,\tau')$ is given by:
\begin{equation}
D(k;\tau,\tau)=(2\pi)^d\delta^{d}(k+k')\frac{1}{(k^{2}+m_{0}^{2})}\biggl(1-\exp\left(-2\tau
(k^{2}+m_{0}^{2})\right)\biggr). \label{44}
\end{equation}
where we assume $\tau=\tau'$. Therefore, for
$\tau\rightarrow\infty$ we recover the Euclidean two-point
function.

The self-interacting theory is beyond the scope of this paper,
however it can be carried out in a straightforward way. In the
next section we present a modification of the Langevin equation
that allows us to treat systems with complex Euclidean actions.
Although non-trivial, it is intuitively obvious that the method
can be extended to interacting field theory.

\section{Stochastic quantization for complex actions}

As an application of the ideas discussed previously, in this
section we show how it is possible to quantize a theory with a
complex action using a non-Markovian Langevin equation. We will be
following similar steps as in Ref. \cite{menezes2}. Consider the
following Euclidean action:
\begin{equation}
S= \frac{1}{2}\int\,d^dx\,\varphi\, K \varphi, \label{sit4}
\end{equation}
with the following Markovian Langevin equation:
\begin{equation}
\frac{\partial}{\partial\tau}\varphi(\tau,x) = -K\varphi +
\eta(\tau,x). \label{sit5}
\end{equation}
where $K$ is an elliptic operator (with some minor changes, our
proof in this section can be made to hyperbolic operators). The
function $\varphi$ is a scalar field, for simplicity, but we can
generalize our results to fields of higher spin. If we let $K$ to
be non-Hermitian, the action in Eq. (\ref{sit4}) becomes complex.
There are many approaches in the literature to deal with complex
actions; one of them is to employ a complex Langevin equation,
separating the field in a real part and in a imaginary part,
$Re(\varphi) = \varphi_1$ and $Im(\varphi) = \varphi_2$
\cite{damre}. With this approach, we get two Langevin equations,
one for each of the two fields $\varphi_1$ and $\varphi_2$.
Another one is to use a modified Langevin equation:
\begin{equation}
\frac{\partial}{\partial\tau}\varphi(\tau,x) =-\int
d^dy\,\kappa(x,y)\frac{\delta\,S_0}{\delta\,\varphi(y)}|_{\varphi(y)=\varphi(\tau,\,y)}+
\eta(\tau,x), \label{231}
\end{equation}
where a subsequent change in the second moment of the noise field
is
\begin{equation}
\langle\, \eta(\tau,x)\,\eta(\tau',x')\,
\rangle_{\eta}\,=2\delta(\tau-\tau')\,\kappa(x,x'). \label{29b}
\end{equation}
With these modifications, we may choose an appropriate kernel:
\begin{equation}
\kappa(x,x') = K_x^{\dag}\delta(x-x'),
\end{equation}
so the Langevin equation becomes
\begin{equation}
\frac{\partial}{\partial\tau}\varphi(\tau,x) = -K.K^{\dag}\varphi +
\eta(\tau,x). \label{sit6}
\end{equation}
We see that we get a ``bosonized" version of the Langevin equation
given by Eq. (\ref{sit5}) and the problem with the convergence
towards an equilibrium disappears, since $K.K^{\dag}$ is a
Hermitian operator. This is the prescription usually employed in
the literature to deal with the stochastic quantization of
fermions \cite{f1} \cite{f2} \cite{f3}. In fact, the root of this
problem lies in the fact that there exists no classical analogue
of fermion fields.

Another approach can be used as well. Let us consider the
following non-Markovian Langevin equation:
\begin{equation}
\frac{\partial}{\partial\tau}\varphi(\tau,x)
=-\int_{0}^{\tau}ds\,M_\Lambda(\tau-s)
\frac{\delta\,S}{\delta\,\varphi(x)}|_{\varphi(x)=\varphi(s,\,x)}+
\eta(\tau,x), \label{93}
\end{equation}
where $M_\Lambda$ is a memory kernel and the stochastic random
field $\eta(\tau,x)$ satisfies the modified Einstein's relations
\begin{equation}
\langle\,\eta(\tau,x)\,\eta(\tau',x')\,\rangle
_{\eta}=2M_{\Lambda}(|\tau-\tau'|)\,\delta^{d}(x-x'). \label{94}
\end{equation}
In this case where $M_{\Lambda}(|\tau-\tau'|)$ has a width in the
fictitious time, the description is Gaussian in spite of being
non-Markovian. For the case of Euclidean free scalar field theory
we have that the generalized Langevin equation reads
\begin{equation}
\frac{\partial}{\partial\tau}\varphi(\tau,x)
=-\int_{0}^{\tau}ds\,M_\Lambda(\tau-s) K\varphi(s,x)+ \eta(\tau,x).
\label{95}
\end{equation}
We shall prove in this section that this method leads to convergence
towards equilibrium, even though we have a complex Langevin
equation.

We can introduce a mode decomposition such as
\begin{equation}
\varphi(\tau,x) = \int\,d\tilde{\mu}(n)\varphi_{n}(\tau)u_{n}(x)
\label{216}
\end{equation}
and
\begin{equation}
\eta(\tau,x) = \int\,d\tilde{\mu}(n)\eta_{n}(\tau)u_{n}(x),
\label{217}
\end{equation}
where the measure $\tilde{\mu}(k)$ depends on the metric we are
interested in. Each Fourier coefficient $\varphi_n$ obeys a
(non-Markovian) Langevin equation given by
\begin{equation}
\frac{\partial}{\partial\tau}\varphi_n(\tau)=
-\lambda_n\int_{0}^{\tau}ds\,M_{\Lambda}(\tau-s) \varphi_n(s)+
\eta_n(\tau), \label{96}
\end{equation}
where $\lambda_n$ is an eigenvalue of the operator $K$ and
$\eta_n(\tau)$ obeys
\begin{equation}
\langle\,\eta_n(\tau)\,\eta_{n'}(\tau')\,\rangle
_{\eta}=2M_{\Lambda}(|\tau-\tau'|)\,\delta^{d}(n,n'). \label{sit94}
\end{equation}
Following Fox \cite{fox1} \cite{fox2}, we define the Laplace
transform of the memory kernel:
\begin{equation}
{M}(z)=\int_{0}^{\infty}d\tau\,M_{\Lambda}(\tau)\,e^{-z\tau}.
\label{n7}
\end{equation}
With the initial condition $\varphi_n(\tau)|_{\tau=0}=0$, the
solution of the Eq. (\ref{96}) reads:
\begin{equation}
\varphi_n(\tau)= \int_{0}^{\infty}d\tau'\,G_n(\tau-\tau')\,
\eta_n(\tau'), \label{k2}
\end{equation}
where using the step function $\theta(\tau)$, the Green function
$G_n(\tau)$ is defined by:
\begin{equation}
G_n(\tau)\equiv\Omega_n(\tau)\,\theta(\tau). \label{k3a}
\end{equation}
The $\Omega_n(\tau)$ function that appears in Eq. (\ref{k3a}) is
defined through its Laplace transform:
\begin{equation}
\Omega_n(\tau)=\biggl(z+\lambda_n{M}(z)\biggr)^{-1}.  \label{k5}
\end{equation}
From Eq. (\ref{k2}) and the modified Einstein relations, we get
that the free scalar correlation function $D_n(\tau,\tau')$ is
given by:
\begin{eqnarray}
&&D_n(\tau,\tau')=\nonumber\\
&& = 2\delta^{d}(n,n')\int_{0}^{\infty} ds\int_{0}^{\infty}
ds'\,G_n(\tau-s)\,G_n(\tau'-s')\,M_\Lambda(\mid s-s' \mid) \nonumber\\
&& = 2\delta^{d}(n,n')\int_{0}^{\tau} ds\int_{0}^{\tau'}
ds'\,\Omega_n(\tau-s)\,\Omega_n(\tau'-s')\,M_\Lambda(\mid
s-s' \mid). \label{k4}
\end{eqnarray}
To proceed we have to write $D_n(\tau,\tau')$ in a simplified way.
Note that the double Laplace transform of the right hand side is
given by:
\begin{eqnarray}
&&\int_{0}^{\infty}d\tau\,e^{-z\tau}\int_{0}^{\infty}d\tau'\,e^{-z\tau'}\int_{0}^{\tau}
ds\int_{0}^{\tau'}
ds'\,\Omega_n(\tau-s)\,\Omega_n(\tau'-s')\,M_\Lambda(\mid
s-s' \mid)=\nonumber\\
&& = \Omega_n(z)\,\Omega_n(z') \int_{0}^{\infty}
ds\int_{0}^{\infty} ds'\,e^{-z's'}e^{-zs}\,M_\Lambda(\mid s-s'
\mid). \label{k6}
\end{eqnarray}
Now, with simple manipulations, we get:
\begin{equation}
\int_{0}^{\infty} ds\int_{0}^{\infty}
ds'\,e^{-z's'}e^{-zs}M_\Lambda(\mid s-s'
\mid)=\frac{{M}(z)+{M}(z')}{z+z'}. \label{k7}
\end{equation}
Therefore, we get the identity:
\begin{eqnarray}
&&\int_{0}^{\infty}d\tau\,e^{-z\tau}\int_{0}^{\infty}d\tau'\,e^{-z\tau'}\int_{0}^{\tau}
ds\int_{0}^{\tau'}
ds'\Omega_n(\tau-s)\,\Omega_n(\tau'-s')\,M_\Lambda(\mid s-s' \mid)=\nonumber\\
&& =
\Omega_n(z)\,\Omega_n(z')\Biggl(\frac{M(z)+M(z')}{z+z'}\Biggr).
\label{k8}
\end{eqnarray}
Remembering Eq. (\ref{k5}), we can show that:
\begin{equation}
\Omega_n(z)\,\Omega_n(z')\Biggl(\frac{M(z)+
M(z')}{z+z'}\Biggr)=\frac{1}{\lambda_n}
\Biggl(\frac{\Omega_n(z)+\Omega_n(z')}{z+z'}-\Omega_n(z)\,\Omega_n(z')\Biggr).
\label{k9}
\end{equation}
So, in parallel with result given by Eq. (\ref{k7}), we finally
obtain a very simple expression for $D_n(\tau,\tau')$ in terms of
$\Omega_n(\tau)$. We have
\begin{equation}
D_n(\tau,\tau')= \frac{2}{\lambda_n}\,\delta^{d}(n,n')\,
\biggl(\Omega_n(\mid
\tau-\tau'\mid)-\Omega_n(\tau)\,\Omega_n(\tau')\biggr).
\label{k10}
\end{equation}
Now, we need an expression for the memory kernel in order to
investigate the convergence of Eq. (\ref{k10}). A series of
kernels were proposed in the literature:
\begin{equation}
M^{(m)}_{\Lambda}(\tau)=\frac{1}{2m!}\Lambda^2(\Lambda^2\mid\tau\mid)^m
\exp\bigl(-\Lambda^2\mid\tau\mid\bigr).      \label{k1}
\end{equation}
For simplicity, we shall take the case for $m=0$. Then, from Eq.
(\ref{n7}), Eq. (\ref{k5}) and Eq. (\ref{k1}), and applying the
inverse Laplace transform \cite{erdelyi}, we obtain the following
expression for the $\Omega$-function:
\begin{equation}
\Omega_n(\tau)=\Biggl(\frac{\Lambda^2}{\beta}\sinh\biggl
(\frac{\beta\tau}{2}\biggr)+\cosh\biggl(\frac{\beta\tau}{2}\biggr)\Biggr)\,
\exp\biggl(-\tau\frac{\,\Lambda^2}{2}\biggr),   \label{k12}
\end{equation}
where we have defined a quantity $\beta$ given by:
\begin{equation}
\beta\equiv x + i y, \label{k13}
\end{equation}
where
\begin{equation}
x = \sqrt{\frac{\Lambda^4 + \alpha_R + |z|}{2}},
\end{equation}
\begin{equation}
y = \alpha_I \sqrt{\frac{1}{2(\Lambda^4 + \alpha_R + |z|)}},
\end{equation}
and, finally:
\begin{equation}
|z| = \sqrt{(\Lambda^4 + \alpha_R)^2 + \alpha_I^2},
\end{equation}
with $\alpha = - 2\Lambda^2\lambda_n$ and we have written $\alpha$
as $\alpha = \alpha_R + i\alpha_I$, where $\alpha_R$ and
$\alpha_I$ are real quantities. Similarly, we have, for the Green
function:
\begin{equation}
G_n(\tau)=\Biggl(\frac{\Lambda^2}{\beta}\sinh\biggl(\frac{\beta\tau}{2}\biggr)
+\cosh\biggl(\frac{\beta\tau}{2}\biggr)\Biggr)\,
\exp\biggl(-\tau\frac{\,\Lambda^2}{2}\biggr)\theta(\tau).
\label{k14}
\end{equation}
Now, in order to have convergence, we must demand that
$G_n\rightarrow 0$ when $\tau\rightarrow\infty$. Using hyperbolic
identities we see that, in order to have $G_n\rightarrow 0$ when
$\tau\rightarrow\infty$, we shall have:
\begin{equation}
 |\lambda_n^R| > \frac{(\lambda_n^I)^2}{2\Lambda^2}, \label{cond1}
\end{equation}
where we have written the eigenvalues as $\lambda_n = \lambda_n^R
+ i\lambda_n^I$. Since $\Lambda$ is, in principle, arbitrary, we
see that the condition given by Eq. (\ref{cond1}) does not seem to
pose any serious restrictions on the eigenvalues of the operator
$K$. However, we have another restriction. Note that, with this
prescription, the function $M(x-y;\tau)$ defined in the Ref.
\cite{menezes2} as
\begin{equation}
M(x-y,\tau)\equiv\int_{0}^{\tau}\,ds\,M_{\Lambda}(\tau-s)\,G(\tau-s,x-y),
\end{equation}
where $G(\tau-s,x)$ is the retarded Green function for the
diffusion problem, whose Fourier transform is given by:
\begin{eqnarray}
M(k,\tau)&=&\frac{\,\Lambda^{2}}{2}\,\frac{1}{9\,\Lambda^{4}
+\beta^{2}}\biggl\{8\,\Lambda^{2}-\nonumber\\
&& 4\,\exp\Bigl(-\frac{3\Lambda^{2}}{2}\tau\Bigr)
\biggl[2\Lambda^{2}\cosh\Bigl(\frac{\beta\tau}{2}\Bigr)+
\biggl(\frac{3\Lambda^{4}}{2\beta}-\frac{\beta}{2}\biggr)
\sinh\Bigl(\frac{\beta\tau}{2}\Bigr)\biggr]\biggr\},
\label{g}
\end{eqnarray}
will no longer be Hermitian. But, if we impose the following
restriction:
\begin{equation}
|\lambda_n^R| < 5 \Lambda^2, \label{cond2}
\end{equation}
its real part will remain positive, which implies that the real
part of the eingenvalues of the Fokker-Planck Hamiltonian defined
therein are positive, assuring, therefore, convergence to
equilibrium, i.e., the system reaches its ground state. It seems
that Eq. (\ref{cond1}) and Eq. (\ref{cond2}) can be imposed
simultaneously without restrictions on $\Lambda$. In particular,
we are allowed to take arbitrarily large values for $\Lambda$,
which would imply that our approach works for almost any value for
the eigenvalues $\lambda_n$. It can be easily proved that the
presence of the zero modes for massless theories destroys the
convergence to equilibrium.

From the results above, it is easy to see that the free two-point
function is given by:
\begin{eqnarray}
&&D_n(\tau,\tau')=\nonumber\\
&& = \frac{2}{\lambda_n}\,\delta^{d}(n,n')\,
\Biggl[\Biggl(\frac{\Lambda^2}{\beta}\,\sinh\biggl(\frac{\beta(\tau-\tau')}{2}\biggr)
+\,\cosh\biggl(\frac{\beta(\tau-\tau')}{2}\biggr)\Biggr)
\,\exp\biggl(-\frac{\,\Lambda^2}{2}\mid(\tau-\tau')\mid\biggr)  \nonumber\\
&&-\Biggl(\frac{\Lambda^2}{\beta}\,\sinh\biggl(\frac{\beta\tau}{2}\biggr)+\,
\cosh\biggl(\frac{\beta\tau}{2}\biggr)\Biggr)\,
\Biggl(\frac{\Lambda^2}{\beta}\,\sinh\biggl(\frac{\beta\tau'}{2}\biggr)+\,
\cosh\biggl(\frac{\beta\tau'}{2}\biggr)\Biggr)\,
\exp\biggl(-\frac{\,\Lambda^2}{2}(\tau+\tau')\biggr)\Biggr].
\label{k15}
\end{eqnarray}
For $\tau=\tau'$, we get:
\begin{equation}
D_n(\tau,\tau)= \frac{2}{\lambda_n}\, \delta^{d}(n,n')\,
\Biggl(1-\Biggl(\frac{\Lambda^2}{\beta}\,\sinh\biggl(\frac{\beta\tau}{2}\biggr)+\,
\cosh\biggl(\frac{\beta\tau}{2}\biggr)\Biggr)^2\exp\bigl(-\Lambda^2\,\tau\bigr)\Biggr)\,
. \label{k16}
\end{equation}
So, in the limit $\tau\rightarrow\infty$, we obtain the following
result:
\begin{equation}
D_n(\tau,\tau)= \frac{2}{\lambda_n}\,\delta^{d}(n,n'). \label{k17}
\end{equation}

Therefore, apart from an unimportant constant, we have obtained
that, in the asymptotic limit $\tau\rightarrow\infty$, we have
reached convergence towards the expected equilibrium, and the
two-point Schwinger function was obtained.

\section{Conclusions and perspectives}

\quad {\,\,\,}There are many examples where the Euclidean field
theory is defined for an imaginary action. Since in this case the
path integral weight is not positive definite, the stochastic
quantization in this situation is problematic. Parisi and Klauder
proposed complex Langevin equations \cite{con4} \cite{con5}, and
some problems of this approach are the following. First of all,
complex Langevin simulations do not converge to a stationary
distribution in many situations. Besides, if it does, it may
converge to many different stationary distributions. The complex
Langevin equation also appears when the original method proposed
by Parisi and Wu is extended to include theories with fermions
\cite{f1} \cite{f2} \cite{f3}. The first question that appears in
this context is if make sense the Brownian problem with
anticommutating numbers. It can be shown that, for massless
fermionic fields, there will not be a convergence factor after
integrating the Markovian Langevin equation. Therefore the
equilibrium is not reached. One way of avoiding this problem is to
introduce a kernel in the Langevin equation describing the
evolution of two Grassmannian fields.

In this paper, we have used the method of the stochastic
quantization to study systems with complex valued path integral
weights. We assumed a Langevin equation with a memory kernel and
Einstein's relation with colored noise. The equilibrium solution
of such Langevin equation was analyzed. We have shown that for a
large class of elliptic non-Hermitean operators which define
different models in quantum field theory converges in the
asymptotic limit of the Markov parameter $\tau\rightarrow\infty$,
and we have obtained the Schwinger functions of the theory.
Although non-trivial, the method proposed can be extended to
interacting field theory with complex actions. The generalization
of the method for this situation is under investigation by the
authors.

\section{Acknowlegements}

This paper was supported by Conselho Nacional de Desenvolvimento
Cientifico e Tecnol{\'o}gico do Brazil (CNPq).

\end{document}